\documentclass[prd,twocolumn,showpacs,preprintnumbers,amsmath,amssymb]{revtex4}

\usepackage{amsmath}
\usepackage{graphicx}
\usepackage{xspace}
\usepackage{color}
\usepackage{url}
\usepackage[utf8]{inputenc}
\usepackage{epstopdf}
\usepackage{hyperref}

\newcommand{\bea}{\begin{eqnarray}}
\newcommand{\eea}{\end{eqnarray}}

\newcommand{\eq}[1]{Eq.~\eqref{#1}}

\newcommand{\GeV}{\,\text{GeV}}
\newcommand{\TeV}{\,\text{TeV}}

\newcommand{\bctaunu}{$b\to c\tau\nu$ }

\begin{document}
%%%%%%%%%%%%%%%%%%%%%%%%%%%%%%%%%
\preprint{PSI-PR-18-07, ZU-TH 23/18}

\title{Importance of Loop Effects in Explaining the Accumulated Evidence for \\New Physics in $B$ Decays with a Vector Leptoquark}

\author{Andreas Crivellin}
\email{andreas.crivellin@cern.ch}
\affiliation{Paul Scherrer Institut, CH--5232 Villigen PSI, Switzerland}

\author{Christoph Greub}
\email{greub@itp.unibe.ch}
\affiliation{Albert Einstein Center for Fundamental Physics, Institute
	for Theoretical Physics,\\ University of Bern, CH-3012 Bern,
	Switzerland}

\author{Dario M\"uller}
\email{dario.mueller@psi.ch}
\affiliation{Paul Scherrer Institut, CH--5232 Villigen PSI, Switzerland}
\affiliation{Physik-Institut, Universit\"at Z\"urich,
	Winterthurerstrasse 190, CH-8057 Z\"urich, Switzerland}

\author{Francesco Saturnino}
\email{saturnino@itp.unibe.ch}
\affiliation{Albert Einstein Center for Fundamental Physics, Institute
	for Theoretical Physics,\\ University of Bern, CH-3012 Bern,
	Switzerland}

%%%%%%%%%%%%%%%%%%%%%%%%%%%%%%%%%%
%%%%%%%%%%%%%%%%%%%%%%%%%%%%%%%%%%
\begin{abstract}
In recent years experiments revealed intriguing hints for new physics (NP) in $B$ decays involving \bctaunu and $b\to s\ell^+\ell^-$ transitions at the $4\,\sigma$ and $5\,\sigma$ level, respectively. In addition, there are slight disagreements in $b\to u \tau\nu$ and  $b\to d\mu^+\mu^-$ observables. While not significant on their own, they point in the same direction. Furthermore, $V_{us}$ extracted from $\tau$ decays shows a slight tension ($\approx2.5\,\sigma$) with its value determined from CKM unitarity and an analysis of BELLE data found an excess in  $B_d\to\tau^+\tau^-$. Concerning NP explanations, the vector leptoquark $SU(2)$ singlet is of special interest since it is the only single particle extension of the Standard Model which can (in principle) address all the anomalies described above. For this purpose, large couplings to $\tau$ leptons are necessary and loop effects, which we calculate herein, become important. Including them in our phenomenological analysis, we find that neither the tension in $V_{us}$ nor the excess in $B_d\to\tau^+\tau^-$ can be fully explained without violating bounds from $K\to\pi\nu\bar\nu$. However, one can account for $b\to c\tau\nu$ and $b\to u\tau\nu$ data finding intriguing correlations with $B_{q}\to\tau^+\tau^-$ and $K\to \pi\nu\bar\nu$. Furthermore, the explanation of $b\to c\tau\nu$ predicts a positive shift in $C_7$ and a negative one in $C_9$, being nicely in agreement with the global fit to $b\to s\ell^+\ell^-$ data. Finally, we point out that one can fully account for \bctaunu and $b\to s\ell^+\ell^-$ without violating bounds from $\tau\to \phi\mu$, $\Upsilon\to\tau\mu$ or $b\to s\tau\mu$ processes.
\end{abstract}


%%%%%%%%%%%%%%%%%%%%%%%%%%%%%%%%%
%%%%%%%%%%%%%%%%%%%%%%%%%%%%%%%%%
\maketitle

\section{Introduction}

So far, the LHC has not directly observed any particles beyond the Standard Model (SM). However, intriguing hints for lepton flavor universality (LFU) violating NP have been acquired:
\vspace{0.2cm}\\
\begin{boldmath}$b\to s(d)\ell^+\ell^-$: \end{boldmath}

The ratios
\begin{align}
	R(K^{(*)})=\frac{{\rm Br}[B\to K^{(*)}\mu^+\mu^-]}{{\rm Br}[B\to K^{(*)} e^+e^-]}\,,
\end{align}
\cite{Aaij:2014ora}(\cite{Aaij:2017vbb}) indicate LFU violation with a combined significance of $\approx4\,\sigma$~\cite{Altmannshofer:2017yso,DAmico:2017mtc,Ciuchini:2017mik,Hiller:2017bzc,Geng:2017svp,Hurth:2017hxg}. Taking also into account all other $b\to s\mu^+\mu^-$ observables, {like the angular observable $P_5^\prime$~\cite{Aaij:2015oid} in the decay $B\to K^*\mu^{+}\mu^{-}$ }, the global fit {of the Wilson coefficients to all available data} even shows compelling evidence~\cite{Capdevila:2017bsm} for NP ($>5\,\sigma$).  

		\begin{figure*}[t]
	\centering
	\begin{minipage}{.25\textwidth}
		$\vcenter{\includegraphics[width=\textwidth]{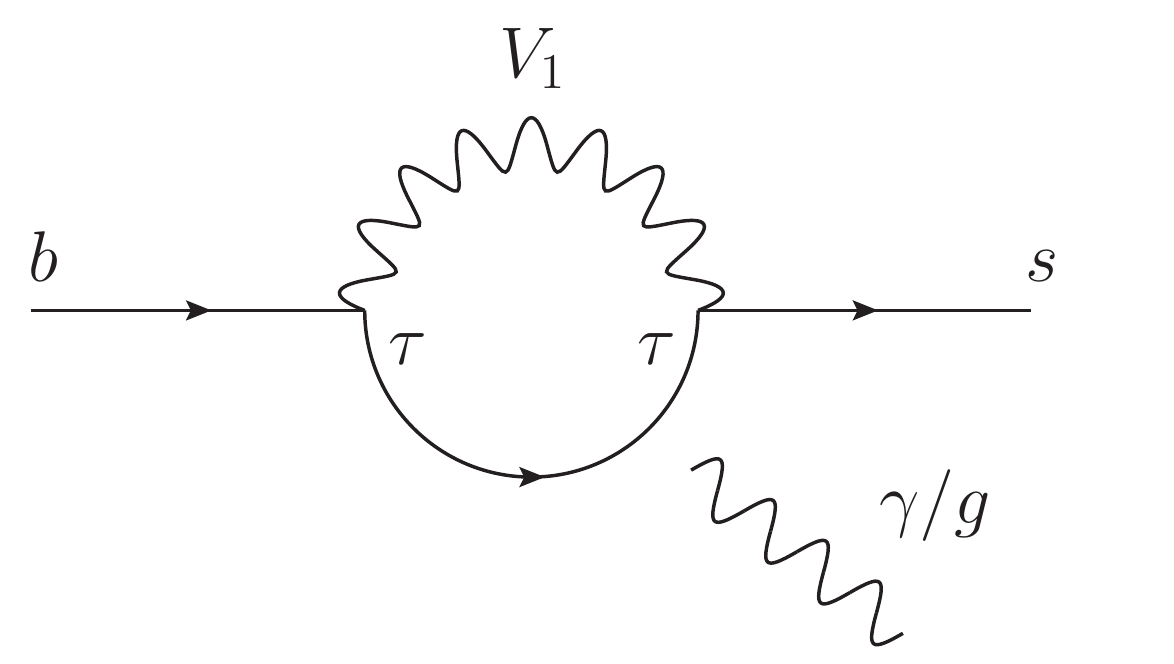}\!\!\!\!}$
	\end{minipage}
	\begin{minipage}{.25\textwidth}
		$\vcenter{\includegraphics[width=\textwidth]{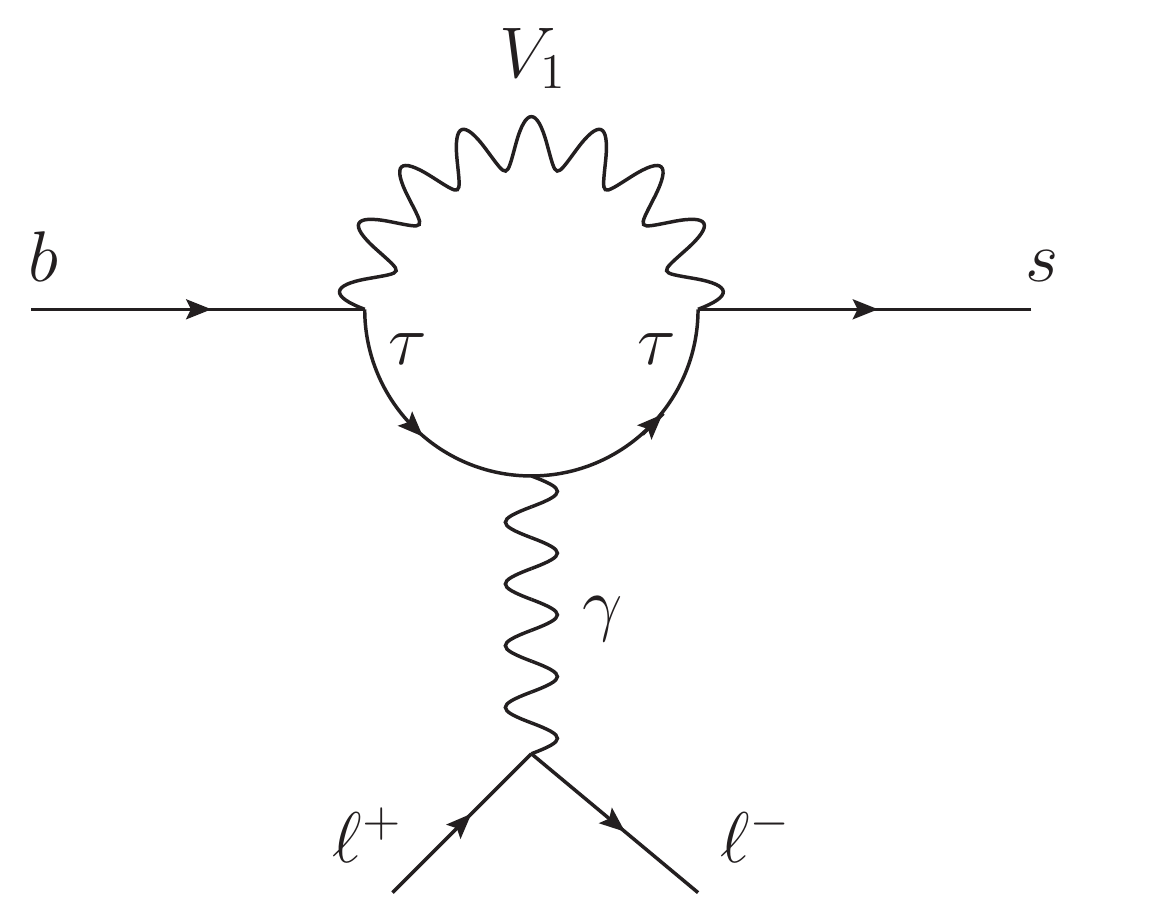} \!\!\!\!}$
	\end{minipage}
	\begin{minipage}{.25\textwidth}
		$\vcenter{\includegraphics[width=\textwidth]{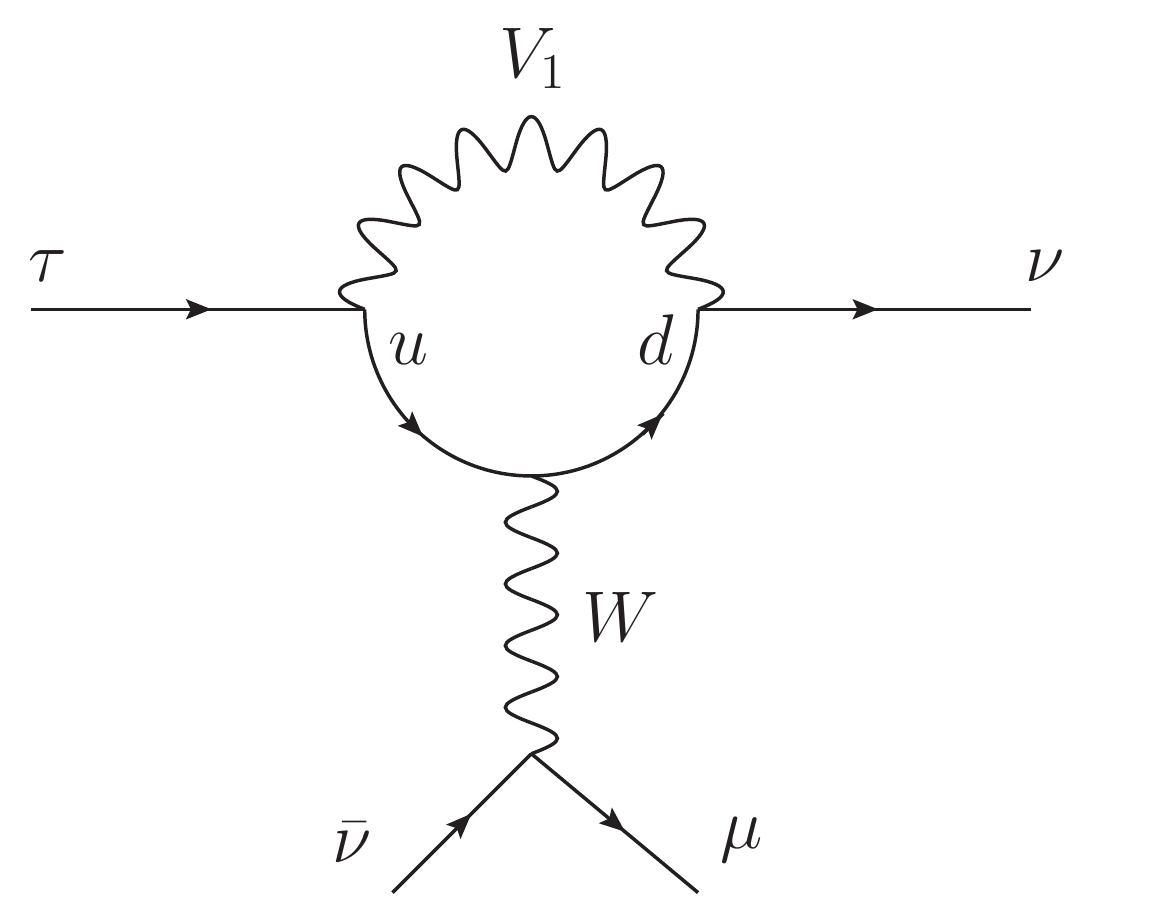}\!\!\!\!}$
	\end{minipage}
	\begin{minipage}{.23\textwidth}
		$\vcenter{\includegraphics[width=\textwidth]{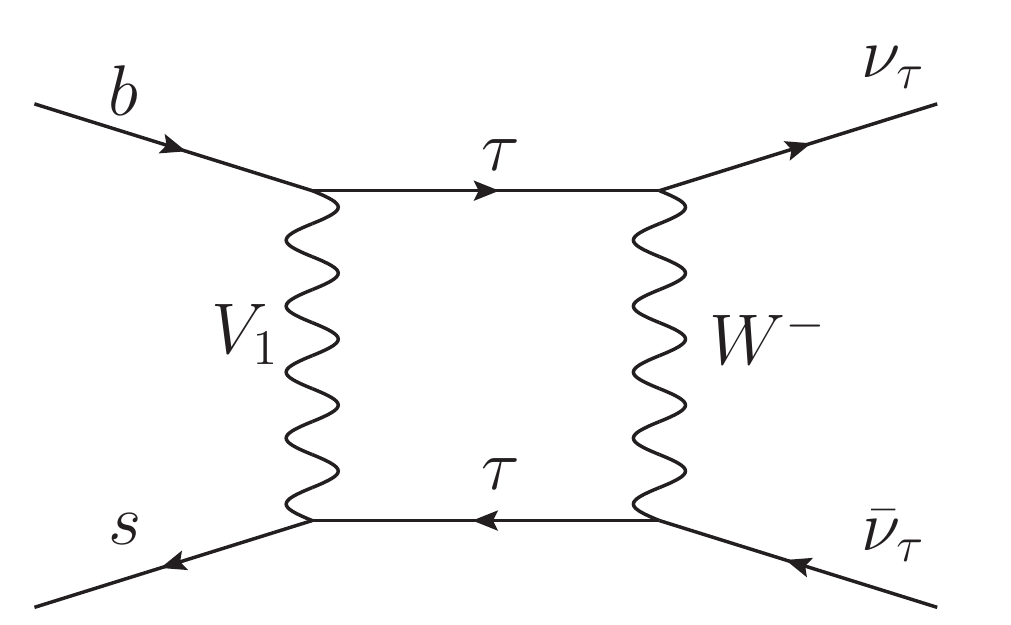}}$
	\end{minipage}
	\caption{Feynman diagrams depicting the one-loop contributions of the vector LQ singlet to $C^{sb}_{7/8}$, $b\to s \ell^+\ell^-$, $\tau\to\mu\nu\bar{\nu}$ and $b\to s \nu \bar{\nu}$ (from left to right).}
	\label{feynman_diagrams}
\end{figure*}

Concerning $b\to d\ell^+\ell^-$ transitions, the theoretical analysis of Ref.~\cite{Hambrock:2015wka} shows that the LHCb measurement of $B\to \pi\mu^+\mu^-$~\cite{Aaij:2015nea} slightly differs from the theory expectation. Even though this is not significant on its own, the central value is very well in agreement with the expectation from $b\to s\ell^+\ell^-$ under the assumption of a $V_{td}/V_{ts}$-like scaling of the NP effect\footnote{{Here, $V$ refers to to the Cabibbo-Kobayashi-Maskawa (CKM) matrix.}}. In other words, an effect of the same order and sign as in $b\to s\ell^+\ell^-$, relative to the SM, is preferred. Furthermore, an (unpublished) analysis of BELLE data found an excess in $B_d\to \tau^+\tau^-$~\cite{Ziegler:2016mtm}.
\vspace{0.2cm}\\
\begin{boldmath}$b\to c(u)\tau\nu$:\end{boldmath}

The ratios
\begin{align} 
R(D^{(*)})=\frac{{\rm Br}[B\to D^{(*)}\tau\nu]}{{\rm Br}[B\to D^{(*)} \ell\nu]}~~\text{with}~~\ell=\{e,\mu\}\,,
\end{align}
which measure LFU violation in the charged current by comparing $\tau$ modes with light leptons ($\ell=e,\mu$), differ in combination from their SM predictions by $\approx4\,\sigma$~\cite{Amhis:2016xyh}. Also, the ratio 
\begin{align}
R(J/\psi)=\frac{{\rm Br}[B_c\to J/\psi\tau\nu]}{{\rm Br}[B_c\to J/\psi\mu\nu]}
\end{align}
\cite{Aaij:2017tyk} exceeds the SM prediction in agreement with the expectations from $R(D^{(*)})$~\cite{Watanabe:2017mip,Chauhan:2017uil}.

Concerning $b\to u\tau\nu$ transitions, the theory prediction for $B\to\tau\nu$ crucially depends on $V_{ub}$. While previous lattice calculations resulted in rather small values of $V_{ub}$, recent calculations give a larger value (see Ref.~\cite{Ricciardi:2016pmh} for an overview). However, the measurement is still above the SM prediction by more than 1$\,\sigma$, as can be 
seen from the global fit~\cite{Charles:2004jd}. In 
\begin{align}
R(\pi)=\frac{{\rm Br}[B\to\pi\tau\nu]}{{\rm Br}[B\to\pi\ell\nu]}
\end{align}
there is also a small disagreement between theory~\cite{Bernlochner:2015mya} and experiment~\cite{Hamer:2015jsa} which does not depend on $V_{ub}$. These results are not significant on their own but lie again above the SM predictions like in the case of $b\to c\tau\nu$.
\vspace{0.2cm}\\	
\begin{boldmath}$V_{us}^\tau$:\end{boldmath} 

$V_{us}$ extracted from $\tau$ lepton decays ($V_{us}^\tau$) shows a tension of $2.5\,\sigma$ compared to the value of $V_{us}$ determined from CKM unitarity ($V_{us}^\text{uni}$)~\cite{Amhis:2016xyh,Lusiani:2017spn}. 	
 
\medskip
The only possible single particle explanation, which can (at least in principle) address all these anomalies is the vector leptoquark (LQ) $SU(2)_L$ singlet $V_1$ with hypercharge{\footnote{In our conventions, the left-handed lepton doublet has hypercharge $-1$.}} $-4/3$\,\cite{Alonso:2015sja,Calibbi:2015kma,Fajfer:2015ycq,Hiller:2016kry,Bhattacharya:2016mcc,Buttazzo:2017ixm,Kumar:2018kmr} arising in the famous Pati-Salam model~\cite{Pati:1974yy}: This LQ can explain $b\to c \tau\nu$ data without violating bounds from  $b\to s\nu\bar{\nu}$ and/or direct searches, provides (at tree level) a {left-handed} solution to $b\to s\ell^+\ell^-$ data, and does not lead to proton decay. Therefore, a sizable effect in $b\to u \tau\nu$ and $b\to d\ell^+\ell^-$ is straightforward, and also an explanation of $V_{us}^{\tau}$ could be possible. A huge enhancement of $b\to s\tau^+\tau^-$ rates is predicted as well~\cite{Capdevila:2017iqn}, making an amplification of $B_d\to\tau^+\tau^-$ possible.

Several attempts to construct a UV completion for this LQ to address the anomalies have been made~\cite{Barbieri:2015yvd,Barbieri:2016las,Assad:2017iib,DiLuzio:2017vat,Calibbi:2017qbu,Bordone:2017bld,Barbieri:2017tuq,Blanke:2018sro,Greljo:2018tuh,Bordone:2018nbg,Matsuzaki:2018jui}. In order to fully account for the $b\to c\tau\nu$ data (while respecting perturbativity), one needs sizable couplings to third generation leptons and $V_1$ generates, via $SU(2)_L$ invariance, also large {contributions to} {the operators $d_id_j\tau\tau$ and $u_iu_j\nu_\tau\nu_\tau$ at tree level. These operators give rise to couplings of down quarks to neutrinos or light charged} leptons at loop level (see Fig.~\ref{feynman_diagrams}).

In this article we will calculate these loop effects \footnote{Similar loop effects for scalar LQs have been calculated in Refs.~\cite{Bobeth:2017ecx,Fajfer:2018bfj,Earl:2018snx}.}, which turn out to be not only numerically important but also give rise to additional correlations among observables. Even though a theory with a massive vector boson without an explicit Higgs sector is not renormalizable, we still identify several phenomenologically important loop effects which are gauge independent and finite and can therefore be calculated reliably (in analogy to flavor observables within the SM). 

%This article is structured as follows: In the next section we will establish our conventions for the various observables. In Sec.~\ref{1loop} we present our results for the loop effects. Sec.~\ref{pheno} is devoted to a phenomenological analysis before we conclude in Sec.~\ref{conclusions}.

\section{Model and One-Loop effects}
\label{1loop}

We work in a simplified model extending the SM by a vector LQ $SU(2)_L$ singlet with hypercharge $-4/3$, mass $M$ and interactions with fermions determined by
\begin{eqnarray}
\mathcal{L}_{V^\mu}=\left(\kappa_{fi}^{L}\overline{Q_f}\gamma_{\mu}L_{i}+\kappa_{fi}^{R}
\overline{d_f}\gamma_{\mu}e_{i}\right)V_{1}^{\mu\dagger}+h.c.\,.
\end{eqnarray}
Here, $Q$ ($L$) are quark (lepton) $SU(2)_L$ doublets, $d$ ($e$) are down quark (charged lepton) singlets and $f,i$ are flavor indices. In the following, we will neglect the right-handed couplings, which are not necessary to explain the anomalies.
% (except for $a_\mu$). 
This then generates the effective four-fermion interactions encoded in 
\begin{equation}
{\mathcal{L}_{{\rm{eff}}}} =- \frac{{{\kappa^{L}_{il}}\kappa^{L*}_{jk}}}{{{M^2}}}\bar Q_i^\alpha {\gamma ^\mu }Q_j^\beta \bar L_k^\beta {\gamma _\mu }L_l^\alpha \,, 
\label{4-fermion}
\end{equation}
where $\alpha$ and $\beta$ label the $SU(2)$ components. After EW symmetry breaking, we work in the down basis; i.e., no CKM elements appear in flavor changing neutral currents of down quarks. We recall our definitions and the tree-level results in the appendix.

In our setup, one-loop effects involving the LQ and third generation leptons ($\tau$'s and $\tau$ neutrinos) can be very important, since we aim for large effects in $b\to c(u)\tau\nu$ and $b\to s(d)\tau^+\tau^-$ processes. In principle, a massive vector boson, like our LQ, without a Higgs sector is not renormalizable. However, in flavor physics most effects can still be calculated reliably since they are gauge independent and finite (also in unitary gauge)\footnote{In this article we followed two approaches to check the results. First, we calculated the results in unitary gauge. Then, we derived the couplings of the LQ Goldstones to SM fermions by requiring the tree-level amplitude to be gauge independent. Finally, we calculated its contribution in $R_\xi$ gauge.}. This is in analogy to the SM, where the contribution of the $W$ to flavor observables can be correctly calculated in unitary gauge without taking into account the Higgs sector. 

We are only interested in effects which are always absent at tree level (like $b\to s\nu\bar{\nu}$ processes) or are not present at tree level due to a specific coupling structure  (like $b\to s\mu^+\mu^-$ processes in the absence of muon couplings). Furthermore, we neglect tiny dimension-8 effects of the SM Higgs particle. In these cases the loop effects are the leading contributions. We calculate all diagrams at leading order in the external momenta using asymptotic expansion~\cite{Smirnov:1994tg}.\\
\vspace{-1.2cm}
\begin{boldmath}
\subsection{$W$ boxes contributing to $d_i\to d_f \nu\bar\nu$}
\end{boldmath}

\begin{figure}[t]
	\begin{center}
		\begin{tabular}{cp{7mm}c}
			\includegraphics[width=0.45\textwidth]{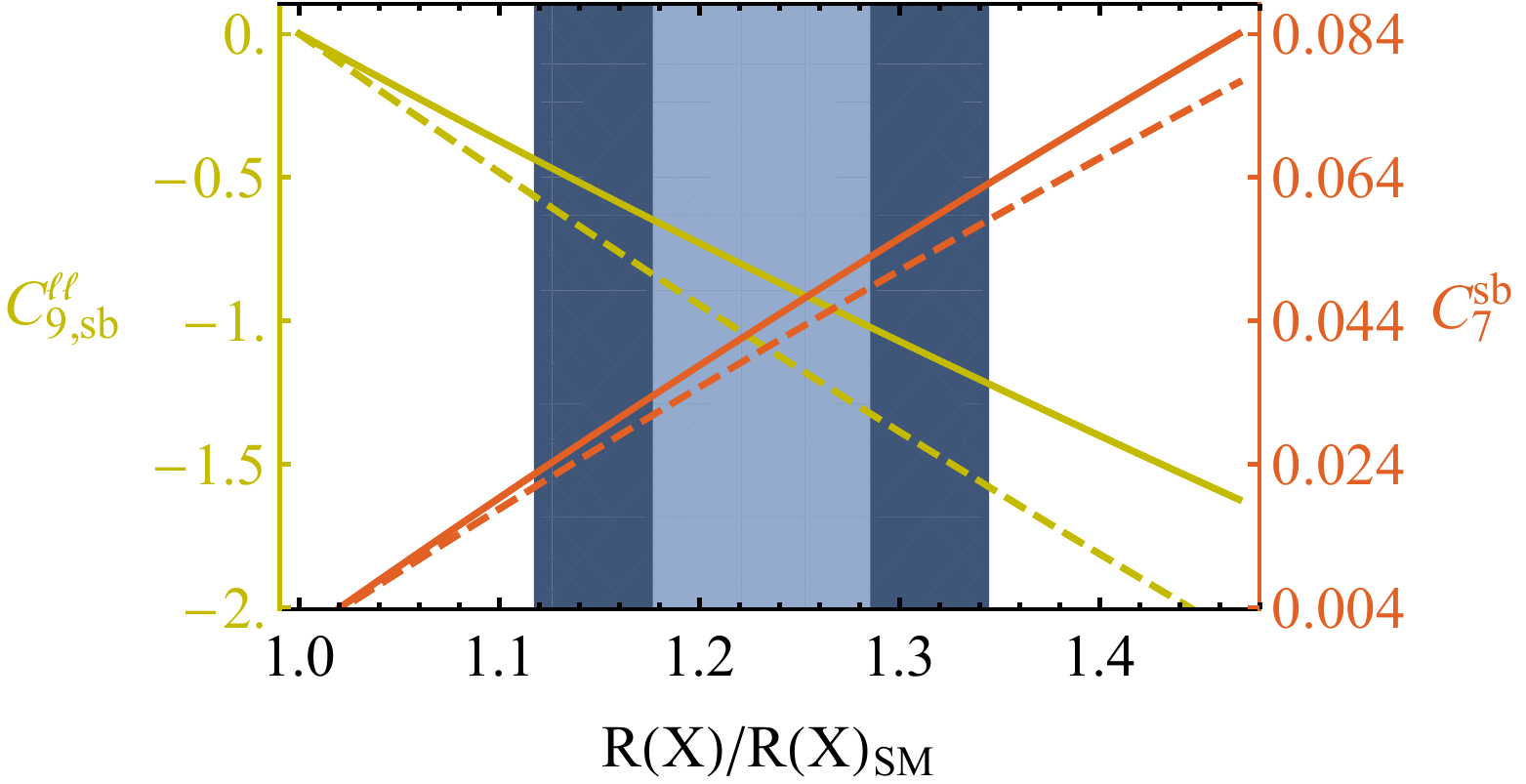}
		\end{tabular}
	\end{center}
	\caption{$C_{9,sb}^{\ell\ell}$ and $C_7^{sb}(\mu_b)$ as functions of $R(X)/R(X)_{\rm SM}$ with $X=\{D,D^{*},J/\psi \}$. The solid lines correspond to $M=1\,$TeV and the dashed ones to $M=5\,$TeV while the (dark) blue region is preferred by $b\to c\tau\nu$ data at the $1\,\sigma$ ($2\,\sigma$) level. From the global fit, taking into account only lepton flavor conserving observables we have $-1.29<C_{9,sb}^{\ell\ell}<-0.87$~\cite{Descotes-Genon:2015uva} and $-0.01<C_7^{sb}(\mu_b)<0.05$~\cite{Capdevila:2017bsm} at the $1\,\sigma$ level. {Therefore, our model predicts just the right sign and size of the effect in $C_{9,sb}^{\ell\ell}$ and $C_7^{sb}(\mu_b)$ necessary to explain $b\to s\ell^+\ell^-$ data, assuming an explanation of $b\to c\tau\nu$.} \label{RDRDSM}}         
\end{figure}

We use the effective Hamiltonian
\begin{align}
\mathcal{H}_{{\rm eff}}^{\nu\nu}&=-\dfrac{4G_F}{\sqrt{2}}V_{td_k}V_{td_j}^{*}\left(C^{fi}_{L,jk}\mathcal{O}_{L,jk}^{fi}+C^{fi}_{R,jk}\mathcal{O}_{R,jk}^{fi}\right)\,,\nonumber\\
\mathcal{O}_{L,jk}^{fi}&=\frac{\alpha}{4\pi}\left[\bar{d}_{j}\gamma^{\mu}P_{L}d_{k}\right] \left[\bar{\nu}_{f}\gamma_{\mu}\left(1-\gamma_5\right)\nu_i\right]\,,\\
\mathcal{O}_{R,jk}^{fi}&=\frac{\alpha}{4\pi}\left[\bar{d}_{j}\gamma^{\mu}P_{R}d_{k}\right] \left[\bar{\nu}_{f}\gamma_{\mu}\left(1-\gamma_5\right)\nu_i\right]\,,\nonumber
\end{align}
with $P_{R(L)}=(1+(-)\gamma_5)/2$ and $G_F$ ($\alpha$) being the Fermi (electromagnetic fine structure) constant. The result of the box contributions involving a $W$ to $d_i\to d_f \nu\bar\nu$ (an example diagram is shown on the right-hand side of Fig.~\ref{feynman_diagrams}) is gauge invariant in $R_\xi$ gauge and the same finite result is obtained in unitary gauge {(with {$e=\sqrt{4\pi\alpha}$} and $m_t$ ($m_W$) the top quark ($W$ boson) mass)}
\begin{widetext}
	\begin{equation}
	C_{L,fa}^{ij}\! =\! \frac{{ - m_W^2}}{{2{e^2}{V_{3a}}V_{3f}^*{M^2}}}\!\!\left( \!\!{6\kappa _{fj}^L\kappa _{ai}^{L*}\log\! \left(\! {\frac{{m_W^2}}{{{M^2}}}} \!\right) \!+\! 3\left( {{V_{3a}}V_{3k}^*\kappa _{ki}^{L*}\kappa _{fj}^L + V_{3f}^*{V_{3k}}\kappa _{kj}^L\kappa _{ai}^{L*}} \right)\!\frac{{\log\! \left( \!{\frac{{m_t^2}}{{m_W^2}}} \!\right)}}{{ {1 - \frac{m_W^2}{m_t^2}}}}\! +\! V_{3f}^*{V_{3k}}\kappa _{kj}^L{V_{3a}}V_{3l}^*\kappa _{li}^{L*}\frac{{m_t^2}}{{m_W^2}}} \!\!\right)
	\label{bsnunu1loop}
	\end{equation}
\end{widetext}

\begin{boldmath}
\subsection{$W$ off-shell penguins contributing to $\tau \to \mu \nu \bar\nu$}
\end{boldmath}

Here (see third diagram in Fig.~\ref{feynman_diagrams}) we obtain again a finite and gauge independent result for the Wilson coefficient; following the analysis of~\cite{Feruglio:2017rjo}, we use
\begin{equation}
{\cal H}_{{\rm{eff}}}^{\tau \mu \nu_f \nu_i } = \frac{{4{G_F}}}{{\sqrt 2 }}D_{L,fi}^{\tau \mu }\left[ {{{\bar \nu }_f}{\gamma ^\sigma }{P_L}{\nu _i}} \right]\left[ {\bar \mu {\gamma _\sigma }{P_L}\tau {\mkern 1mu} } \right]\,,
\end{equation}
with
\begin{equation}
D_{L,fi}^{\tau \mu } = {N_c}{\delta _{i2}}\frac{{V_{3k}^*\kappa _{kf}^{L*}\kappa _{33}^L}}{{32{\pi ^2}}}\frac{{m_t^2}}{{{M^2}}}\left(\! {1 + 2\log \left( {\frac{{m_t^2}}{{{M^2}}}} \right)} \!\right){\mkern 1mu}.
\end{equation}
We find, in agreement with Ref.~\cite{Buttazzo:2017ixm}, that the effect is small.
\vspace{0.2cm}
\begin{boldmath}
\subsection{Photon and gluon penguins}
\end{boldmath}

We use the standard Hamiltonian (see, for example, Ref.~\cite{Descotes-Genon:2015uva}) also defined in the appendix. For on-shell photons and gluons the result of the left-hand diagram in Fig.~\ref{feynman_diagrams} is finite in unitary gauge and the same result is obtained in $R_\xi$ gauge:
\begin{eqnarray}
C^{sb}_{7(8)} &=& \frac{-\sqrt{2}}{{{G_F}{V_{tb}}V_{ts}^*{M^2}}}\frac{{11}}{{72}}\left(\frac{{5}}{{48}}\right){\kappa^{L} _{2i}\kappa _{3i}^{L*}}\,.
\end{eqnarray}
Taking into account the running from the LQ scale $\mu_{LQ}=M=1 \TeV$ down to $\mu_{b}=5 \GeV$ (see, e.g., Refs.~\cite{Borzumati:1998tg,Borzumati:1999qt}), we obtain
\begin{eqnarray}
C_{7}^{sb}(\mu_{b})\approx 0.29~{\kappa^{L} _{2i}\kappa _{3i}^{L*}}\,.\label{C7low}
%C_{8}^{sb}(\mu_{b})=\left(0.185+0.004i\right)\sum\limits_j {\kappa^{L} _{2j}\kappa _{3j}^{L*}}\,.
\end{eqnarray}

\begin{figure}[t!]
	\begin{center}
		\begin{tabular}{cp{7mm}c}
			\includegraphics[width=0.5\textwidth]{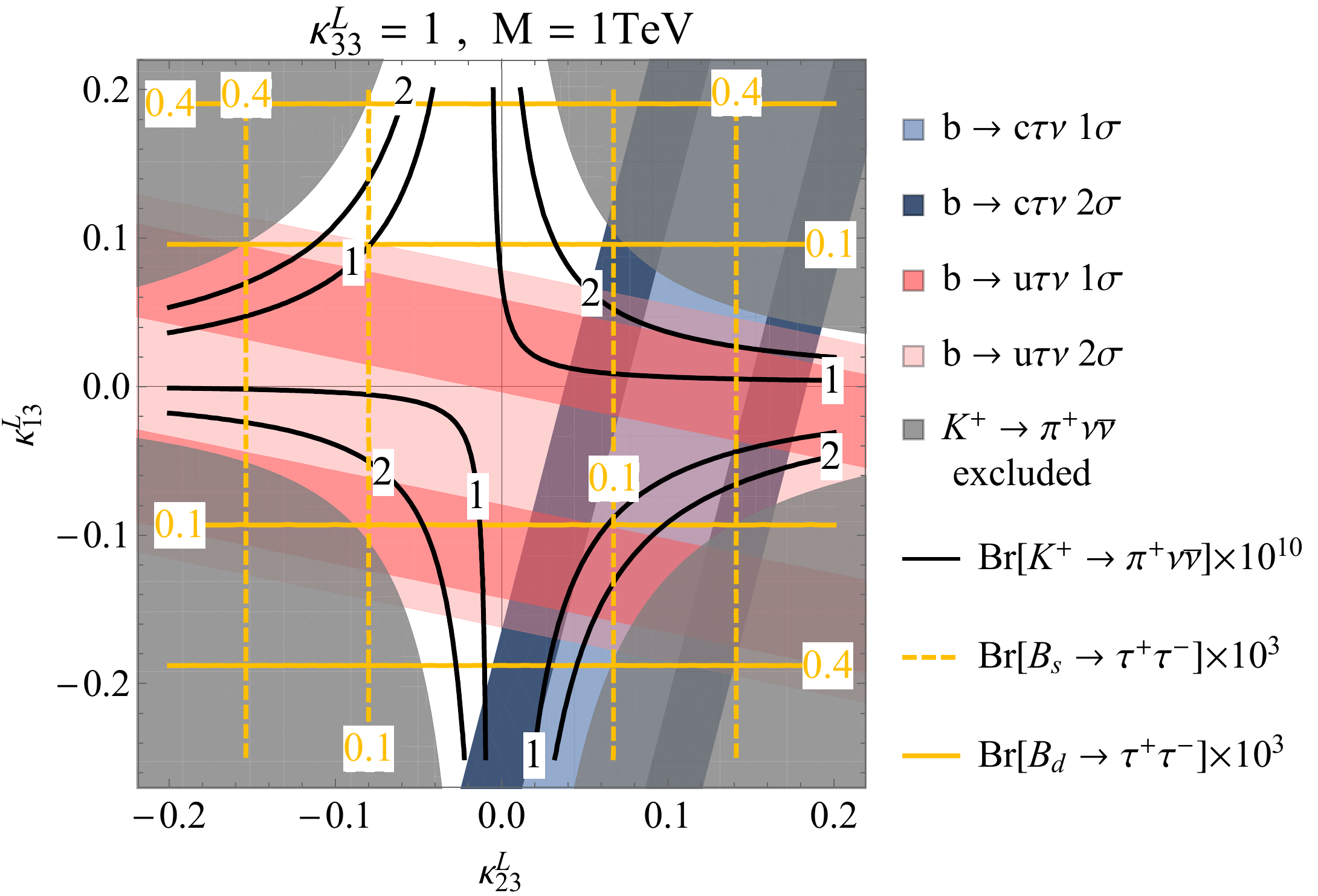}
		\end{tabular}
	\end{center}
	\caption{Predictions for $B_q\to\tau^+\tau^-$ and $K\to\pi\nu\bar{\nu}$ (contour lines) in the $\kappa^{L}_{13}$-$\kappa^{L}_{23}$ plane for $M=1\,{\rm TeV}$ and $\kappa^{L}_{33}=1$. The colored regions are preferred by $b\to c(u)\tau\nu$ data, where we naively averaged {(i.e., we computed the weighted average of the observables and added their errors in quadrature, disregarding correlations)} $R(D^{(*)})$ and $R(J/\psi)$ or $R(\pi)$ and $B\to\tau\nu$, respectively. The gray region is excluded by $K^+\to\pi^+\nu\bar\nu$. Here we assumed all couplings $\kappa_{ij}^L$ to be real.}         
	\label{kappa13_kappa23}
\end{figure}

For off-shell photons the full result (second diagram in Fig.~\ref{feynman_diagrams}) for the amplitude is gauge dependent and, in general, divergent. However, one can calculate the mixing of $C_{9,sb}^{\tau\tau}=-C_{10,sb}^{\tau\tau}$ into the four-fermion operators $O_{9,sb}^{\ell\ell}$ (containing light leptons as well) within the effective theory (i.e. after integrating out the LQ at tree level). In this way, a gauge independent result is obtained and the leading logarithm of the (unknown) full result is recovered. For off-shell photons we thus calculate the effect in the EFT (below the LQ scale), generating the following mixing into the four-fermion operators with light leptons:
\begin{eqnarray}
C_{9,sb}^{\ell\ell} &=& \frac{\sqrt 2 }{{{G_F V_{tb}}V_{ts}^*}{M^2}}\frac{1}{{6}}\log \left( {\frac{{{M^2}}}{{\mu_b^2}}} \right){\kappa _{2i}^{L}\kappa _{3i}^{L*}}\,. \label{C9}
\end{eqnarray}
Note that this result is model independent (at leading-log accuracy) in the sense that it does not depend on the model which generates $C_{9,sb}^{\tau\tau}=-C_{10,sb}^{\tau\tau}$. In principle, there are also $Z$ penguins generating $C_{9,sb}^{\ell\ell}$ and $C_{10,sb}^{\ell\ell}$. However, this effect is suppressed by light lepton masses (or small momenta) and is therefore of dimension 8. Further, note that there are no box diagram contributions which generate $\bar s b \bar\mu\mu$ $ (\bar s b\bar ee)$ operators if the couplings of the LQ to muons (electrons) are zero at tree level.
\vspace{2mm}

\begin{boldmath}
\subsection{Box diagrams with LQs}
\end{boldmath}

What cannot be calculated consistently are box diagrams involving only LQs~\cite{Barbieri:2016las}. Here, the results are divergent in unitary gauge which corresponds to a gauge dependence in $R_\xi$ gauge. However, these effects are suppressed if $|\kappa^L|<g_2$ and can be further suppressed in the presence of vectorlike fermions by a GIM-like mechanism~\cite{Calibbi:2017qbu}~which, in analogy to the SM, would render the result finite. 

\begin{figure}[t]
	\begin{center}
		\begin{tabular}{cp{7mm}c}
			\includegraphics[width=0.5\textwidth]{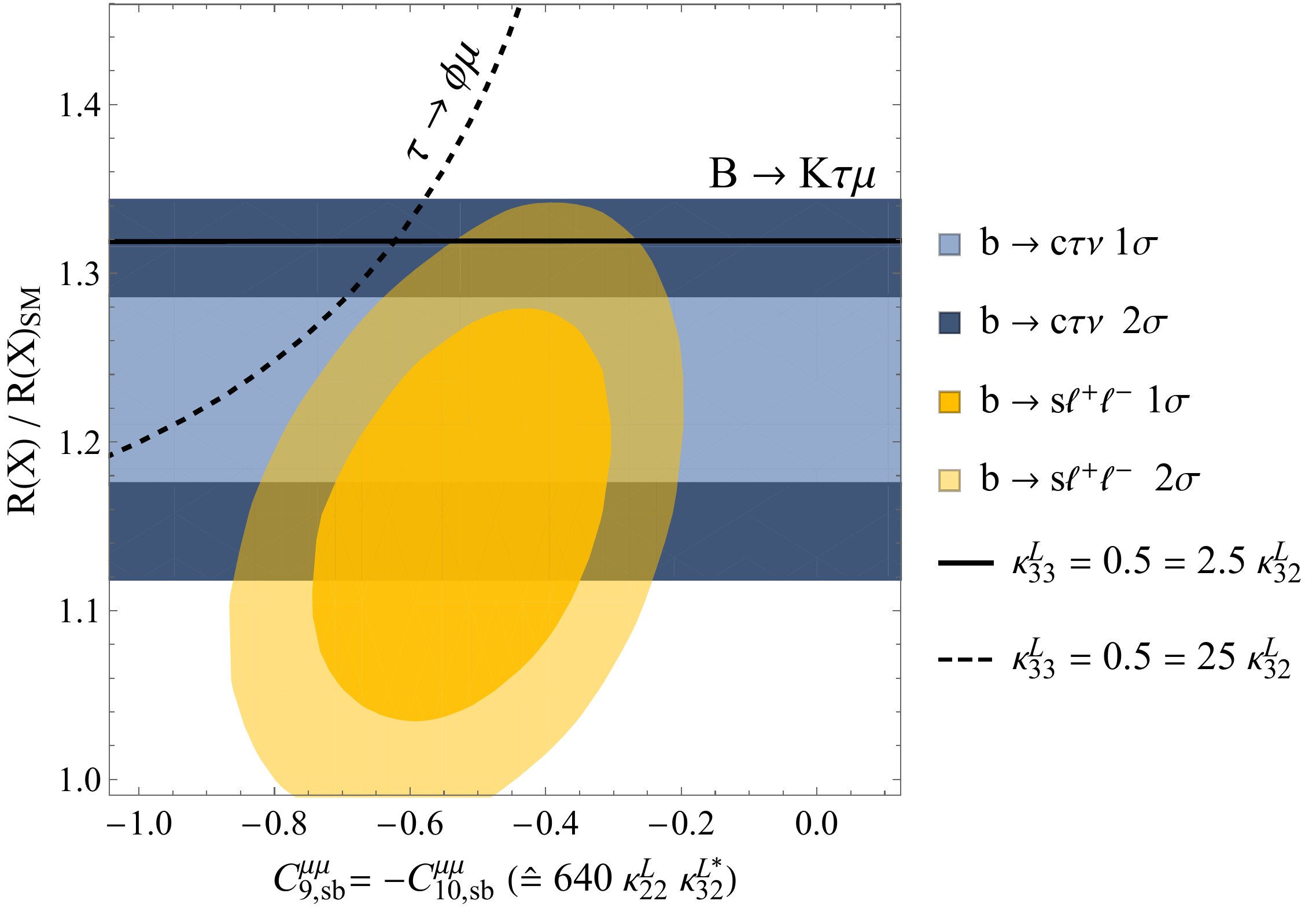}
		\end{tabular}
	\end{center}
	\caption{Allowed (colored) regions in the $C_{9,sb}^{\mu\mu}\!=\!-C_{10,sb}^{\mu\mu}$  $\left(\widehat{=}\;640 \kappa_{22}^L\kappa_{32}^{L*}\right)$ -- $R(X)/R(X)_{\rm SM}$ plane for $M=1\,$ TeV and $X=D,D^*,J/\psi$ at the $1\,\sigma$ and $2\,\sigma$ level for $\kappa^L_{33}V_{cb}\ll\kappa^L_{23}$. The region above the black dashed (solid) line is excluded by $\tau\to\phi\mu$ ($B\to K\tau\mu$) for $\kappa^L_{33}=0.5=25\kappa_{32}^L$ ($\kappa^L_{33}=0.5=2.5\kappa_{32}^L$). The bound from $\tau\to\phi\mu$ ($B\to K\tau\mu$) depends on $\kappa_{33}^L$ and $\kappa_{32}^L$ and gets stronger if $\kappa_{32}^L$ gets smaller (larger). That is, for $\kappa^L_{33}=0.5$ and $2.7\lessapprox\kappa_{33}^L/\kappa_{32}^L\lessapprox27$, the whole $2\,\sigma$ region preferred by $b\to c\tau\nu$ and $b\to s\ell^+\ell^-$ data is consistent with $B\to K\tau\mu$ and $\tau\to\phi\mu$.}         
	\label{C9-R(D)}
\end{figure}

\vspace{0.6cm}
\section{Phenomenology}
\label{pheno}

Assuming $\kappa^L_{33} V_{cb}\ll\kappa^L_{23}$, one is safe from LHC bounds, and the effects in $B_s\to\tau^+\tau^-$, $C_7^{sb}(\mu_b)$ (\eq{C7low}) and $C_{9,sb}^{\ell\ell}$ (\eq{C9}) directly depend on $R(X)/R(X)_{\rm SM}$ (with $X=D^{(*)},J/\psi$). In Fig.~\ref{RDRDSM} we show these dependences. Intriguingly, the effect generated in $C_7^{sb}(\mu_b)$ and $C_{9,sb}^{\ell\ell}$, within the preferred region from $b\to c\tau\nu$ data, exactly overlaps with the $1\,\sigma$ ranges
of the model independent fit to $b\to s\mu^+\mu^-$ data excluding LFU violating observables~\cite{Altmannshofer:2014rta,Descotes-Genon:2015uva} (therefore, only $P_5^\prime $ etc. but not $R(K^{(*)})$ can be explained).

Let us now include the effect of $\kappa_{13}^L$.  Here, many correlations arise. First of all, $b\to c(u)\tau\nu$ is already at tree level correlated to $b\to s(d)\tau^+\tau^-$. In addition, the $W$ boxes in \eq{bsnunu1loop} generate effects in $B\to K^{(*)} (\pi)\nu\bar{\nu}$ and $K\to\pi\nu\bar{\nu}$. While the bounds from $B\to K^{(*)} (\pi)\nu\bar{\nu}$ turn out to be weaker than the ones from $B_{q}\to \tau^+\tau^-$, there are striking correlations with $K\to\pi\nu\bar{\nu}$, as can be seen from Fig.~\ref{kappa13_kappa23}. Furthermore, we get an effect
\begin{equation}
\delta V_{us}^{\tau}=\frac{V_{us}^{\tau}-V_{us}^{\tau (0)}}{V_{us}^{\rm uni}}\approx -C_{us}^{\tau\tau}\,,
\end{equation}
where $V_{us}^{\tau (0)}$ is the CKM matrix element extracted from $\tau$ decays without NP. However, \eq{bsnunu1loop} generates $K\to\pi\nu\bar{\nu}$, and respecting these bounds, the relative effect in $V_{us}^\tau$ can only be at the per-mill level, $\left|\delta V_{us}^{\tau}\right|\approx 0.05\% $, excluding the possibility to account for the discrepancy of $|V^{\rm uni}_{us}|=0.22547 \pm 0.00095$ versus $|V^{\tau}_{us}| = 0.2212 \pm 0.0014$ ~\cite{Lusiani:2017spn,Amhis:2016xyh}. The same is true about $B_d\to\tau^+\tau^-$, where the currently preferred region of analysis using BELLE data~\cite{Ziegler:2016mtm} of ${\rm{Br}}\left[{B_d \to \tau^+\tau^-}\right]_{ \rm{exp} } =  \left( 4.39{}^{+ 0.80}_{-0.83} \pm 0.45\right)$ lies outside the plot range.

Now, in addition to the couplings $\kappa^L_{33}$ and $\kappa^L_{23}$, we allow nonvanishing $\kappa_{32}^L$ and $\kappa_{22}^L$. These couplings give rise to tree-level effects in $b\to s\mu^+\mu^-$. In Fig.~\ref{C9-R(D)} we show the allowed (colored) regions from $b\to s\mu^+\mu^-$ and $b\to c\tau\nu$ as well as the exclusions from $b\to s\tau\mu$ and $\tau\to\phi\mu$. Note that a simultaneous explanation of the anomalies is perfectly possible since the colored regions overlap and do not extend to the parameter space excluded by $b\to s\tau\mu$ and $\tau\to\phi\mu$. Interestingly, due to the loop effects originating from the $b\to c\tau\nu$ explanation, we predict a flavor universal effect in $C_{9,sb}^{\ell\ell}$ and $C_7^{sb}$ which is supplemented by a tree-level effect of the form $C_{9,sb}^{\mu\mu}=-C_{10,sb}^{\mu\mu}$ with muons only. This means that the relative NP effect compared to the SM in lepton flavor conserving observables (like $P_5^\prime$) should be larger than in $R(K^{(*)})$, which is in perfect agreement with the global fit\footnote{See Ref.~\cite{Alguero:2018nvb} for a recent analysis of such scenarios.}.
\vspace*{-0.4cm}
\section{Conclusions}
\label{conclusions}
The vector leptoquark $SU(2)$ singlet is a prime NP candidate to explain the current hints for LFU violation. In this article we calculated and studied the important loop effects arising within such a model and performed a phenomenological analysis. We find:
\smallskip

An explanation of $b\to c\tau\nu$ data generates lepton flavor universal effects in $b\to s\ell^{+}\ell^{-}$ transitions which nicely agree with the model independent fit (see Fig.~\ref{RDRDSM}). Therefore, the $C_{9}=-C_{10}$-like tree-level effect, which is in general LFU violating, is supplemented by these effects generating a new pattern for the Wilson coefficients. This can be tested with future data. That is, with more precise measurements of lepton flavour universality violating and lepton flavor universality conserving effects, one can test if in fact there is a lepton flavor universality conserving contribution in addition to the lepton flavor universality violating ones~\cite{Alguero:2018nvb}. Similar conclusions hold for the correlations between $b\to u\tau\nu$ data generating lepton flavor universal effects in $b\to d\ell^{+}\ell^{-}$ processes.

NP in $b\to c(u)\tau\nu$ generates important effects in $B_{s(d)}\to\tau^+\tau^-$ which are even correlated to $b\to s(d)\nu\bar{\nu}$ processes and $K\to\pi\nu\bar{\nu}$ via $W$ box contributions (see right-hand diagram in Fig.~\ref{feynman_diagrams}).
The $V_{us}^\tau$ puzzle (like the CP asymmetry in $\tau \to K_S\pi\nu$~\cite{Cirigliano:2017tqn}) cannot be solved due to the stringent constraints from $K\to\pi\nu\bar{\nu}$, and because of $b\to u\tau\nu$ bounds one cannot fully account for the BELLE excess in $B_d\to\tau^+\tau^-$ (see Fig.~\ref{kappa13_kappa23}).

$b\to c\tau\nu$ and $b\to s\ell^{+}\ell^{-}$ data can be simultaneously explained without violating other bounds like $\tau\to\phi\mu$ (see Fig.~\ref{C9-R(D)}). Furthermore, one could at the same time also account for NP effects in $b\to d\mu^{+}\mu^{-}$ without violating $K_L\to \mu^+\mu^-$ bounds.
\smallskip

{\it Acknowledgements} --- {\small We are very grateful to Joaquim Matias and Bernat Capdevilla for providing us with the fit necessary for the $b\to s\ell^+\ell^-$ region in Fig.~\ref{C9-R(D)} whose work is supported by an explora grant (FPA2014-61478-EXP) and to Aleksey Rusov for providing us with the fit for $C_{9,bd}^{\mu\mu}$. The work of A.C. and D.M. is supported by an Ambizione Grant of the Swiss National Science Foundation (PZ00P2\_154834). The work of C.G. and F.S. is supported by the Swiss National Foundation under Grant 200020\_175449/1.}

\section{Appendix}

In this appendix we recall the tree-level results for the observables and give details on the experimental situation.

\begin{boldmath}
\subsection{$d_k\to d_j\ell_f^-\ell_i^+$}
\end{boldmath}
We define the effective Hamiltonian as
\begin{align}
\begin{split}
\mathcal{H}_{{\rm eff}}^{\ell\ell}&=- \dfrac{ 4 G_F }{\sqrt 2}V_{td_k}V_{td_j}^{*} \sum^{10}\limits_{a =7} C_{a,d_j d_k}^{fi} \mathcal{O}_{a,d_j d_k}^{fi}\,,\\
{\mathcal{O}_{7(8)}^{jk}} &=\dfrac{e (g_s)}{16\pi^2}m_k[\bar d_j{\sigma^{\mu\nu} } (T^a) P_R d_k]F_{\mu\nu}(G^a_{\mu\nu})\,,\\
{\mathcal{O}_{9(10),jk}^{fi}} &=\dfrac{\alpha }{4\pi}[\bar d_j{\gamma ^\mu } P_L d_k]\,[\bar\ell_f{\gamma _\mu }(\gamma_5)\ell_i] \,,
\end{split}
\end{align}
and obtain at tree level
\begin{eqnarray}
C_{9,jk}^{fi}=-C_{10,jk}^{fi}=& \dfrac{{ -\sqrt 2} }{{2{G_F}{V_{td_k}}V_{td_j}^*}}\dfrac{\pi }{\alpha }\dfrac{1}{{{M^2}}}\kappa_{ji}^{L}\kappa_{kf}^{L*}\,.
\label{eq:effHam}
\end{eqnarray}

For $b\to s\mu^+\mu^-$ transitions, the allowed range is~\cite{Capdevila:2017bsm}
\begin{equation}
-0.37 (-0.49) \geq  C_{9,sb}^{\mu\mu}=-C_{10,sb}^{\mu\mu} \geq (-0.75) -0.88\,,
\end{equation}
at the ($1\,\sigma$) $2\,\sigma$ level, assuming a vanishing effect in electrons. In $b\to d\mu^+\mu^-$ transitions one finds for the Wilson coefficients 
\begin{equation}
C_{9,db}^{\mu\mu}=-C_{10,db}^{\mu\mu}= -1.9 \pm 1.1\,,
\end{equation}
assuming them to be real~\cite{Hambrock:2015wka}. For $\tau$ leptons we have experimentally~\cite{Aaij:2017xqt}
\begin{eqnarray}
{\rm{Br}}{\left[{B_s \to {\tau ^ + }{\tau ^ - }} \right]_{{\rm{exp}}}} \le 6.8 \times {10^{ - 3}}\quad(95\%\,\mathrm{C.L.})\,,
\end{eqnarray}
and for $B_d\to\tau^+\tau^-$ there is a (unpublished) measurement of BELLE~\cite{Ziegler:2016mtm} and an upper limit of LHCb~\cite{Aaij:2017xqt}
\begin{align}
\begin{split}
{\rm{Br}}\left[{B_d \to \tau^+\tau^-}\right]^{\rm{BELLE}}_{ \rm{exp} } &=  \left( 4.39{}^{+ 0.80}_{-0.83} \pm 0.45\right) \times 10^{-3}\,,\\
{\rm{Br}}\left[B_d \to \tau^+\tau^-\right]^{\rm{LHCb}}_{\rm{exp}} &\le  2.1\times {10^{ - 3}} \quad(95\%\,\mathrm{C.L.})\,.
\end{split}
\end{align}
Both are compatible at the $2\,\sigma$ level. The SM predictions are given by~\cite{Bobeth:2013uxa,Bobeth:2014tza}
\begin{eqnarray}
\begin{aligned}
{\rm{Br}}{\left[{B_{s} \to {\tau ^ + }{\tau ^ - }} \right]_{{\rm{SM}}}} = \left( {7.73 \pm 0.49} \right) \times {10^{ - 7}}\,,\\
{\rm{Br}}{\left[{B_{d} \to {\tau ^ + }{\tau ^ - }} \right]_{{\rm{SM}}}} = \left( 2.22\pm 0.19 \right)  \times {10^{ - 8}}\,.
\end{aligned}
\end{eqnarray}
In our model, we have
\begin{equation}
\dfrac{{\rm{Br}}\left[{{B_{q}} \to {\tau ^ + }{\tau ^ - }} \right]}{{\rm{Br}}{\left[{{B_{q}} \to {\tau ^ + }{\tau ^ - }} \right]_{\rm SM}}}=\left|{ {1 + \frac{{C_{10,qb}^{\tau\tau}}}{{C_{10,qb}^{\rm SM}}}}}\right|^2\,,
\label{Bstautau}
\end{equation}
with $q=s,d$ and $C_{10,qb}^{\rm SM} \approx -4.3$~\cite{Bobeth:1999mk,Huber:2005ig}. For the analysis of $B\to K^{(*)}\tau\mu$ we will use the results of Ref.~\cite{Crivellin:2015era}.

The short distance contribution to the branching ratio of $K_{L}\to \mu^{+}\mu^{-}$ is given by~\cite{Buras:2015yca} (with the Hamiltonian defined e.g. in Ref.~\cite{Blanke:2008yr})
\begin{equation}
{\rm{Br}}\left[K_{L}\to \mu^{+}\mu^{-}\right]_{\rm{SD}}=a_{L}\left(\frac{{\mathop{\rm Re}\nolimits}\left[\lambda_{t}\tilde{Y}\right]}{\lambda^5}+\frac{{\mathop{\rm Re}\nolimits}\left[\lambda_{c}\right]}{\lambda}P_{c}^{Y}\right)^2\nonumber
\end{equation}
with the numerical input
\begin{eqnarray}
a_{L}&=&2.01\times10^{-9}\,,\;
\tilde{Y}=Y_{SM}-s_{W}^2 C_{10,sd}^{\mu\mu}\,,\nonumber\\
Y_{SM}&=&1.018\left(\frac{m_t}{170\,{\rm{GeV}}}\right)^{1.56}\,,\;
P_{c}^{Y}=0.115\pm0.017\,, \nonumber\\ \lambda_i&=& V_{is}^{*}V_{id}\,,\;
\lambda=\left|V_{us}\right|\,.
\end{eqnarray}
The upper experimental limit for the short distance contribution is~\cite{Isidori:2003ts}
\begin{equation}
{\rm{Br}}\left[K_{L}\to \mu^{+}\mu^{-}\right]_{\rm{SD}}<2.5\times10^{-9}\,.
\end{equation}

Using Ref.\cite{Bhattacharya:2016mcc} we have
\begin{equation}
{\rm{Br}}\left[\tau\!\to\!\phi\mu\right]=\frac{f_{\phi}^{2}m_{\tau}^{3}\tau_{\tau}}{128\pi}\frac{\left|\kappa_{22}^{L}\kappa_{23}^{L*}\right|^2}{M^4}\left(\!\!1-\frac{m_{\phi}^{2}}{m_{\tau}^{2}}\right)^{\!2}\!\!\left(\!\!1+2\frac{m_{\phi}^{2}}{m_{\tau}^{2}}\!\right)\nonumber
\end{equation}
with the current experimental limit~\cite{Miyazaki:2011xe}
\begin{equation}
{\rm{Br}}\left[\tau\to\phi\mu\right]<8.4\times10^{-8} \quad(90\%\,\mathrm{C.L.})\,.
\end{equation}

If we consider $\Upsilon(nS)\to\tau\mu$, we have~\cite{Kumar:2018kmr}
\begin{eqnarray}
{\rm{Br}}\left[\Upsilon(3S)\to\tau\mu\right]&=&2.6\times 10^{-7}\frac{\big|\kappa^{L}_{32}\kappa^{L*}_{33}\big|^2}{M^4(\TeV)}\,.
\end{eqnarray}
Comparing this to the experimental limit ${\rm{Br}}\left[\Upsilon(3S)\to\tau\mu\right]< 3.1\times10^{-6}~(90\%\,\mathrm{C.L.})$ of Ref.~\cite{Lees:2010jk}, this does not pose relevant constraints on our model.
\vspace*{-0.55cm}
\begin{boldmath}
\subsection{$d_k\to d_j\nu_i\bar{\nu}_f$}
\end{boldmath}
We use the conventions
\begin{align}
\begin{split}
\mathcal{H}_{{\rm eff}}^{\nu\nu}&=-\dfrac{4G_F}{\sqrt{2}}V_{td_k}V_{td_j}^{*}\left(C^{fi}_{L,jk}\mathcal{O}_{L,jk}^{fi}+C^{fi}_{R,jk}\mathcal{O}_{R,jk}^{fi}\right)\,,\\
\mathcal{O}_{L(R),jk}^{fi}&=\frac{\alpha}{4\pi}\left[\bar{d}_{j}\gamma^{\mu}P_{L(R)}d_{k}\right] \left[\bar{\nu}_{f}\gamma_{\mu}\left(1-\gamma_5\right)\nu_i\right]\,.
\end{split}
\end{align}
Note that the LQ does not contribute at tree level. 

For $K\to \pi\nu\bar{\nu}$ we use Ref.~\cite{Buras:2004qb} with the updated numerical values given in Ref.~\cite{Buras:2015qea} resulting in 
\begin{eqnarray}
&&\!\!\!\!\!\!\!\!{\rm{Br}}\left[ {{K^ \pm } \to {\pi ^ \pm }\nu \bar \nu } \right] = \frac{1}{3}\left( {1 + {\Delta _{EM}}} \right){\eta _\pm }\times\nonumber\\
&&\!\!\!\!\!\sum\limits_{f,i = 1}^3 \left[\left(\frac{{\mathop{\rm Im}\nolimits} {{\left[ {{\lambda _t}\tilde X_L^{fi}} \right]}}}{\lambda ^5}\right)^{\!\!2} \!\!\!+\! \left( \frac{{\mathop{\rm Re}\nolimits} \left[ {{\lambda _c}} \right]}{\lambda }{P_c}{\delta _{fi}} + \frac{{\mathop{\rm Re}\nolimits} \left[ {{\lambda _t}\tilde X_L^{fi}} \right]}{\lambda ^5} \right)^{\!\!\!2}\right],\nonumber\\
&&{\rm{Br}}\left[ {{K_L} \to \pi \nu \bar \nu } \right] = \frac{1}{3}{\eta _L}\sum\limits_{f,i = 1}^3 \left({\frac{{\mathop{\rm Im}\nolimits} {{\left[ {{\lambda _t}\tilde X_L^{fi}} \right]}}}{\lambda ^5}}\right)^{\!\!2}\,,
\end{eqnarray}
with
\begin{eqnarray}
\begin{aligned}
\tilde X_L^{fi} &= X_{L}^{{\rm{SM}},fi} - {s_W^2}{C_{L,sd}^{fi}}\,,\;
{P_c} = 0.404 \pm 0.024\\
{\eta _\pm } &= \left( {5.173 \pm 0.025} \right){10^{ - 11}}{\left[ {\frac{\lambda }{{0.225}}} \right]^8}\,,\\
{\eta _L} &= \left( {2.231 \pm 0.013} \right){10^{ - 10}}{\left[ {\frac{\lambda }{{0.225}}} \right]^8\,,}\\
{\Delta _{EM}} &=  - 0.003\,,\;
X_{L}^{{\rm{SM}},fi} = \left( {1.481 \pm 0.005 \pm 0.008} \right){\delta _{fi}}\,.
\end{aligned}
\end{eqnarray}

For $B\to K^{(*)}\nu\bar{\nu}$ we follow Ref.~\cite{Buras:2014fpa}, giving $C_{L,sb}^{{\rm SM},fi}\approx-1.47/s_W^2\delta_{fi}$, and the branching ratios normalized by the SM predictions read
\begin{equation}
{R_{K^{(*)}}^{\nu\bar{\nu}}} = 
\frac{1}{3}\sum\limits_{f,i=1}^3 \dfrac{ \big|{C_{L,sb}^{fi}}\big|^2}{\big|{C_{L,sb}^{{\rm SM},fi}}\big|^2} \,.
\end{equation}
This has to be compared to the current experimental limits ${R_K^{\nu\bar{\nu}}} < 3.9$ and ${R_{{K^*}}^{\nu\bar{\nu}}} < 2.7$~\cite{Grygier:2017tzo} (both at $90\%\,\mathrm{C.L.}$). The future BELLE II sensitivity for $B\to K^{(*)}\nu\bar{\nu}$ is 30\% of the SM branching ratio~\cite{Abe:2010gxa}.
\vspace{0.2cm}
\begin{boldmath}
\subsection{$d_k\to u_j\bar{\nu}\ell^{-}$}
\end{boldmath}
We define the effective Hamiltonian as
\begin{equation}
{\mathcal{H}_{{\rm{eff}}}^{\ell_f\nu_i}} = \dfrac{{4{G_F}}}{{\sqrt 2 }}{V_{jk}}C_{jk}^{fi}\left[ {\bar u_{j}{\gamma ^\mu }{P_L}d_k} \right]\left[ {{{\bar \ell }_f}{\gamma _\mu }{P_L}{\nu _i}} \right]\,,
\end{equation}
where in the SM $C_{jk,{\rm SM}}^{fi}=\delta_{fi}$. The contribution of our model is given by
\begin{equation}
C_{jk}^{fi}=\frac{\sqrt{2}}{4G_{F}M^2}\frac{V_{jl}}{V_{jk}}\kappa^{L}_{li}\kappa^{L*}_{kf}\,.
\end{equation}
With these conventions we have for $b\to c\tau\nu$ transitions
\begin{equation}
{R\left( {{X}} \right) \mathord{\left/
		{\vphantom {{R\left( {{X}} \right)} {R{{\left( {{D^{\left( * \right)}}} \right)}_{{\rm{SM}}}}}}} \right.
		\kern-\nulldelimiterspace} {R{{\left( {{X}} \right)}_{{\rm{SM}}}}}} = \sum\limits_{i = 1}^3 {{{\left| {{\delta _{3i}} + C_{cb}^{\tau i}} \right|}^2}}\,,
\end{equation}
with $X=\{D,D^*,J/\psi\}$, assuming vanishing contributions to the muon and electron channels. We obtain the analogous expression for $b\to u\tau\nu$.

Concerning $\tau\to K(\pi)\nu$ we find that the CKM element $V_{us}^\tau$ extracted from these decays is given in terms of the one determined in the absence of NP contributions ($V_{us}^{\tau(0)}$) by 
\begin{align}
V_{us}^\tau=V_{us}^{\tau(0)}/(1+C_{us}^{\tau\tau})\,,
\end{align}
where we neglected LFV effects. This has to be compared to~\cite{Lusiani:2017spn,Amhis:2016xyh}
$|V^{\rm uni}_{us}|=0.22547 \pm 0.00095$ and $|V^{\tau}_{us}| = 0.2212 \pm 0.0014$.

In Ref.~\cite{Bernlochner:2015mya} the analysis gives
\begin{align}
\begin{split}
R\left(\pi\right)_{\rm{exp}}&=1.05\pm0.51\,,\\
R\left(\pi\right)_{\rm{SM}}&=0.641\pm0.016\,.
\end{split}
\end{align}
For $B\to\tau\nu$ we use the PDG value~\cite{Patrignani:2016xqp} and the SM prediction of Ref.~\cite{Charles:2004jd} at the $2\,\sigma$ level
\begin{align}
\begin{split}
{\rm{Br}[B\to\tau\nu]_{\rm exp}}&=\left(1.09\pm0.24\right)\times 10^{-4}\,,\\
{\rm{Br}[B\to\tau\nu]_{\rm SM}}&=\left(0.851^{+0.079}_{-0.077}\right)\times 10^{-4}\, .
\end{split}
\end{align}

\bibliography{BIB}

\end{document}